# Towards effective shaping of large-amplitude high-fidelity Schrödinger cat states by inefficient photon number resolving detection

Sergey A. Podoshvedov[1*], Dmitry A. Kuts[1] and Ba An Nguyen[2,3]

[1]*Laboratory of Quantum Information Processing and Quantum Computing, Institute of Natural and Exact Sciences, South Ural State University (SUSU), Lenin Av. 76, Chelyabinsk, Russia*
[2]*Thang Long Institute of Mathematics and Applied Sciences (TIMAS), Thang Long University, Nghiem Xuan Yem, Hoang Mai, Hanoi, Vietnam*
[3]*Institute of Physics, Vietnam Academy of Science and Technology (VAST), 18 Hoang Quoc Viet, Cau Giay, Hanoi, Vietnam*
[*] sapodo68@gmail.com

**Abstract:** We propose an efficient way to generate optical analogs of both even and odd Schrödinger cat states (SCSs) with high fidelity, large amplitude and reasonable generation rate. The resource consumed are a single-mode squeezed vacuum state (SMSV) and possibly a single photon or nothing. We report the generation of even (odd) SCS with amplitude 4.2, fidelity higher than 0.99 and reasonable generation rate by subtraction of 30 (31) photons from SMSV by ideal photon number detection. In the case of inefficient detectors, maintaining SCS's size and its fidelity at the same level as in the case of ideal detectors results in a dramatic decrease in the success probability. In order to have certain harmony between the three characteristics (large amplitude, high fidelity and acceptable success probability for the generation scheme) in the case of imperfect detection, highly transmitting beam splitters should not be used and number of the subtracted photons must be reduced to 10 (11).

Quantum mechanics involves a number of thought experiments that, in most cases, are used to show its weakness in various interpretations. Schrödinger cat states (SCSs) [1] serve basis for doubts about the lack of a clear boundary between the quantum and everyday classical realms being defining feature of quantum theory. Realization of such bizarre physical objects is expected to resolve the puzzle, at what degree of macroscopicity, if it exists, the object goes on to be quantum [2,3]? In optics, the SCS corresponds to a superposition of coherent states $|+\beta\rangle$ and $|-\beta\rangle$ with complex amplitudes $\pm\beta$. Although each of component coherent states is considered to be most classical [4], their superposition corresponding to SCS is nonclassical. The squared absolute value of amplitude of the component coherent state $|\beta|^2$, which is equal to its mean photon number, is treated as size of the associated SCS. In this work, for simplicity, we assume $\beta$ to be real positive, so the SCS size can be characterized simply by $\beta$. In order to recognize an optical SCS macroscopic object, its size must be at least much larger than the quantum uncertainty $1/\sqrt{2}$ of the position observable in the coherent state [5].

In addition to their fundamental importance, the SCSs have high application potentials in teleportation [6-9], quantum metrology [10], quantum computation [11,12] as well as quantum information processing with hybrid entangled states composed of coherent and photonic states [13-15]. SCSs, also called coherent-state qubits, can be considered to be practical when the coherent components are nearly orthogonal $\langle -\beta|\beta\rangle \approx 0$ what starts from SCS's size $\beta \geq 2$ [13] that is regarded as a significant experimental achievement. Within this context, a number of schemes for generation of optical large-amplitude SCSs in "flying" modes are demonstrated [16-20]. However, even in best experiments (see, e.g., [20]), the obtained amplitude $\beta = 1.85$ of the SCS is not sufficient for full use of the states in further quantum processing. This is mainly due to the fact that the used experimental methods do not allow restoring photon number distribution of SCSs with amplitude larger than 2 that must be shifted towards higher order number states and centered near the Fock state $|n\rangle$ with $n \sim |\beta|^2$. Fidelity is another important characteristic for SCSs to serve as a potential source. It is desirable that the fidelity of the output state will be as high as possible (ideally $\geq 0.99$) in order to be able to efficiently convert coherent states on a balanced beamsplitter like $|\beta\rangle_1|\beta\rangle_2 \to |\sqrt{2}\beta\rangle_1|0\rangle_2$ [7,11-13]. Otherwise (i.e., the fidelity of the output state is low), the output photon number distribution may contain unwanted coincidence measurement events and quantum processing with coherent states may become ineffective. Despite the large number of theoretical proposals for the SCS generation [21-27], implementation of large-amplitude high-fidelity source of event-ready even/odd SCSs has remained a challenging problem even from a theoretical point of view. A standard method for the SCSs generation is photon subtraction [16-27] from the single-mode squeezed vacuum (SMSV) which is a typical nonclassical state containing only even photon numbers (i.e., state with definite, even, parity). In this method, part of the photons resided in the SMSV is diverted to detection channel. The redirection of photons by the beam splitter with a high transmittance coefficient when only a small part of the photons are directed to the measurement mode can significantly reduce the generation rate, which can be defined as $P_{suc} \cdot f_{sus}$ with $f_{os}$ the operating rate of the optical scheme and $P_{suc}$ the SCSs generation probability. This third important facet of SCS source can take daunting small values, especially, in the case of registration of Fock states with a large number of photons in optical setup with high transmission beam splitter. Therefore, heralded methods need improvement to brighten its potential in shaping large-amplitude high-fidelity even/odd SCSs generated with a rate sufficiently high for the subsequent continuous-variable (CV) quantum computing.

Here, we give a total analysis of large-amplitude high-fidelity even/odd SCSs source capable to work at high generation rate under ideal conditions. The driving force behind an efficient SCS source is a highly efficient photon number resolving (PNR) detector [28,29]. We report the generation of large-amplitude high-fidelity even/odd SCSs (say, those with $\beta \simeq 4.2$ which we sometimes call bright cats with fidelity higher than 0.99 and success probability achieving even greater than 2%) by extracting a large number of photons (say, either 30 or 31) from the SMSV by ideal PNR detectors with quantum efficiency $\eta = 1$. No restrictions on the beam splitter, being second most important element after PNR detector, are imposed. Unfortunately, the inefficiency of the detector ($\eta < 1$) can significantly reduce the generation rate, however, leaving the amplitude and fidelity at the same level. To maintain a balance between the three important characteristics, it is necessary to reduce the number of extracted photons (say, to 10 or 11) when dealing with imperfect PNR detectors.

We begin by describing ingredients to the optical scheme in Fig. 1. There are two input modes to the beam splitter $(BS)$, labeled mode 1 and mode 2 in the figure. The input state to mode 1 is a single-mode squeezed vacuum (SMSV) state

$$|SMSV\rangle = \sum_{l=0}^{\infty} b_{2l}|2l\rangle, \qquad (1)$$

with amplitudes

$$b_{2l} = \frac{1}{\sqrt{\cosh s}} \left(\frac{\tanh s}{2}\right)^l \frac{\sqrt{(2l)!}}{l!}, \qquad (2)$$

where $s$ is the squeezing parameter of the SMSV state, whereas the input state to mode 2 may be either the vacuum state $|0\rangle$ or a single-photon state $|1\rangle$. The SMSV state (1) has a definite parity, which is referred to as even, since it consists exclusively of Fock states with an even number of photons. Another state with definite even parity is the even SCS which has the form (1) but the amplitudes differ. Namely, if we denote the even SCS with amplitude $\beta$ by $|SCS_+(\beta)\rangle \equiv |SCS_+\rangle$, then its full expression reads

$$|SCS_+\rangle = 2N_+(\beta) \exp(-\beta^2/2) \sum_{l=0}^{\infty} \frac{\beta^{2l}}{\sqrt{(2l)!}} |2l\rangle, \qquad (3)$$

where $N_+ = \left(2(1 + \exp(-2\beta^2))\right)^{-1/2}$ is the normalization factor. As an example of the state with definite odd parity, we may mention the so-called odd SCS denoted by $|SCS_-(\beta)\rangle \equiv |SCS_-\rangle$ with its full expression as

$$|SCS_-\rangle = 2N_-(\beta) \exp(-\beta^2/2) \sum_{l=0}^{\infty} \frac{\beta^{2l+1}}{\sqrt{(2l+1)!}} |2l + 1\rangle, \qquad (4)$$

with the corresponding normalization factor $N_- = \left(2(1 - \exp(-2\beta^2))\right)^{-1/2}$.

If we subtract (add) an even number of photons from (to) the state (1), then it will naturally preserve its parity (i.e., it will remain even), but it will be transformed to a state with a photon number distribution different from that of the initial one. Likewise, subtraction (addition) of an odd number of photons from (to) the SMSV state results in a state with odd parity with totally different photon number distribution. To make use of such properties, let us first consider possible output states of mode 1 in Fig. 1 when nothing is inputted into mode 2 (formally, it implies that the input state of mode 2 is $|0\rangle$). Then, after the SMSV state passes through a lossless beam splitter $BS$ with $t$ and $r$ the real transmittance and reflectance coefficients satisfying the physical condition $t^2 + r^2 = 1$, we have [30]

$$BS_{12}(|SMSV\rangle_1 |0\rangle_2) = \sum_{l=0}^{\infty} b_{2l} BS_{12}(|2l\rangle_1 |0\rangle_2)$$
$$= \sum_{m=0}^{\infty} \frac{r^{2m}}{\sqrt{(2m)!}} \left(\sum_{k=0}^{\infty} b_{2(k+m)} t^{2k} \sqrt{\frac{(2(k+m))!}{(2k)!}} |2k\rangle_1\right) |2m\rangle_2 -$$
$$\sum_{m=0}^{\infty} \frac{tr^{2m+1}}{\sqrt{(2m+1)!}} \left(\sum_{k=0}^{\infty} b_{2(k+m+1)} t^{2k} \sqrt{\frac{(2(k+m+1))!}{(2k+1)!}} |2k + 1\rangle_1\right) |2m + 1\rangle_2. \qquad (5)$$

Depending on the parity of the measurement outcome in the output mode 2, i.e., whether an even or an odd number of photons is detected by PNR detector [28,29] in Fig. 1, the output state of mode 1 differs. If the detector finds an even photon number $n = 2m$, then the following conditional state is outputted

$$|\Psi_{2m}^{(0)}\rangle = L_{2m} \sum_{k=0}^{\infty} b_{2(k+m)} t^{2k} \sqrt{\frac{(2(m+k))!}{(2k)!}} |2k\rangle, \qquad (6)$$

where the normalization factor is $L_{2m}$ and the success probability is

$$P_{2m}^{(0)} = \frac{(1-t^2)^{2m}}{(2m)! L_{2m}^2}, \qquad (7)$$

with the superindex "(0)" indicating the case when the input mode 2 is the vacuum state $|0\rangle$. Otherwise, if an odd photon number $n = 2m + 1$ is found, then the output state of mode 1 reads

$$|\Psi_{2m+1}^{(0)}\rangle = L_{2m+1} \sum_{k=0}^{\infty} b_{2(k+m+1)} t^{2k} \sqrt{\frac{(2(m+k+1))!}{(2k+1)!}} |2k + 1\rangle, \qquad (8)$$

with the normalization factor $L_{2m+1}$ and the success probability

$$P_{2m+1}^{(0)} = \frac{t^2(1-t^2)^{2m+1}}{(2m+1)! L_{2m+1}^2}. \qquad (9)$$

The physical condition $\sum_{m=0}^{\infty}\left(P_{2m}^{(0)} + P_{2m+1}^{(0)}\right) = 1$ is guaranteed as can be directly checked. Obviously, the conditional state in Eq. (6) is even while that in Eq. (8) is odd. Roundly speaking, the parity of the output state (6) and the original input SMSV state (1) are the same, but their expansion amplitudes (i.e., their photon number distribution) are not. Also for the output state (8), its parity changes compared with that of the original input SMSV state (1) and its distribution in photon numbers is totally different. Moreover, the amplitudes $b_{2k}$ of the original SMSV change to $b_{2(k+m)}(b_{2(k+m+1)})$ due to the redistribution of photons by the BS. The proximity between the output states and the even/odd SCSs is evaluated by the fidelities $F_{2m}^{(0)}(\beta) = \left|\langle SCS_+ | \Psi_{2m}^{(0)} \rangle\right|^2$ and $F_{2m+1}^{(0)}(\beta) = \left|\langle SCS_- | \Psi_{2m+1}^{(0)} \rangle\right|^2$, respectively.

The output states $|\Psi_n^{(0)}\rangle$, with $n = 2m$ or $n = 2m + 1$, depend on the squeezing parameter $s$ of the input squeezed state $|SMSV\rangle$, the transmittance coefficient $t$ of the BS and the detection outcome $n$, while the target states $|SCS_\pm\rangle$ depend only on $\beta$. So, the fidelities $F_n^{(0)}$ depend on all the parameters $s, t, n$ and $\beta$, but the probabilities $P_n^{(0)}$ depend only on $s, t$ and $n$. We display in Figs. 2a and 2b our numerical simulation for the dependence on $\beta$ and $n$ of the fidelity maximalized over $s$ and $t$, i.e., $F_{max}^{(0)} = max_{s,t} F_n^{(0)}(\beta, n, s, t)$. Such maximum values of the fidelity appear as a smooth curve with the following behaviors: for a given $n$ the maximalized fidelity decreases with increasing $\beta$ and for a given $\beta$ it increases with increasing $n$. Generally, an arbitrarily high value of the fidelity with a desired value of $\beta$ can be obtained if $n$ is large enough. For example, as seen from Figs. 2a and 2b, a fidelity greater than or equal to 0.99 for $\beta \geq 2$ is achievable for $n \geq 10$. Subtraction of smaller numbers of photon would lead to the generation of the even/odd SCSs state of amplitude $\beta < 2$ [16-20]. It is also observed that for a given $n$ there is a value of amplitude $\beta_{0.99}^{(0)}(n)$ such that the maximalized fidelity is higher than 0.99 for $\beta < \beta_{0.99}^{(0)}(n)$ but falls down rather quickly for $\beta \geq \beta_{0.99}^{(0)}(n)$. The value of $\beta_{0.99}^{(0)}(n)$ itself grows with increasing $n$: for instance, $\beta_{0.99}^{(0)}(30) = 3.1 > \beta_{0.99}^{(0)}(12) = 2$. Although values of the parameters $(s, t)$ can be determined that make the fidelities maximal, these maximizing parameters are largely scattered, i.e., they appear very different even with a slight variation in $\beta$. This feature leads to a significant spread in the output state's generation probability as shown by colored symbols in Figs. 2c and 2d that correspond to Figs. 2a and 2b, respectively. The probability distributions in Figs. 2c and 2d look like structureless swarms.

To get rid of such structureless swarms of success probabilities, we optimize the experimental parameters. Contour lines of the fidelities for the state $|\Psi_{2m}^{(0)}\rangle$ in Fig. 3a clarify the optimization procedure. So, the high fidelities $F_n^{(0)} \geq 0.99$ occupy a narrow area (highlighted in blue) stretching from top to bottom on the $(s, t)$ plane. Choosing values of $(s, t)$ from a given curved stripe provides fidelity $F_n^{(0)} \geq 0.99$ but the corresponding probabilities in Eq. (7) differ very much. Within this stripe, one can find values of $(s, t)$ denoted by $(s, t)_{Fid}$ that provide the highest possible fidelity shown in Figs. 2a and 2b. Yet, within the same stripe one can also find values of $(s, t)$ denoted by $(s, t)_{Prob}$ that provide the maximum success probabilities. These values do not match $(s, t)_{Fid} \neq (s, t)_{Prob}$. Visually, the difference is shown by two dots in Fig.3a: one in red for $(s, t)_{Fid}$ and the other in black for $(s, t)_{Prob}$. Finally, if we are interested only in obtaining maximum fidelities, then we have curves in Figs. 2a and 2c. If we are interested in a more practical case with both sufficiently high fidelity and highest possible success probability, then we refer to Figs. 2e and 2h. Note that fidelities $F_{2m}^{(0)} \geq 0.99$ are also observed for small values of the squeezing amplitude $s$ in the case of a high-transmission beam splitter (Fig. 3a), which, nevertheless, can sharply reduce the success probability which is not practical. Concerning generation of an odd SCS, analogical curved stripe corresponding to the fidelity $F_{2m+1}^{(0)} \geq 0.99$ is also observed. The plots in Figs. 2b and 2d are made for those $(s,t)_{Fid}$ that provide maximum fidelity, while Figs. 2g and 2i correspond to $(s,t)_{Prob}$ with maximum success probability and fidelity greater than 0.99. The optimization procedure over probability allows one to observe a more regular pattern in Figs. 2h and 2i which exhibit the following properties. For a given $\beta$ the optimized probability decreases with increasing $n$ no matter $n$ is even or odd. However, the dependence of the optimized probability on $\beta$ is sensitive to both the value of $\beta$ itself and the parity of $n$: when $\beta$ is small, e.g., $\beta < 2$, it decreases (increases) with increasing $\beta$ for a given even (odd) $n$, but when $\beta$ is large, e.g. $\beta \geq 2$, it saturates to a certain $n$-dependent value despite of the further increase in $\beta$.

In order to maintain a high fidelity for a larger value of $\beta$, consider the case when a single photon is inputted into mode 2 of the scheme in Fig. 1. Then, the mixing of the SMSV state and the single photon on the BS results in [30]

$$BS_{12}(|SMSV\rangle_1|1\rangle_2) = BS_{12}(\sum_{l=0}^{\infty} b_{2l}|2l\rangle_1|1\rangle_2) = \sum_{l=0}^{\infty} b_{2l}BS_{12}(|2l\rangle_1|1\rangle_2) =$$
$$r\left(\sum_{k=0}^{\infty} b_{2k} t^{2k}\sqrt{2k+1}|2k+1\rangle_1\right)|0\rangle_2 -$$
$$t^2 \sum_{m=1}^{\infty} \frac{r^{2m-1}\sqrt{(2m)!}}{(2m-1)!} \left(\sum_{k=0}^{\infty} b_{2(k+m)} t^{2k} \sqrt{\frac{(2(k+m))!}{(2k+1)!}} \left(1 - \frac{2k+1}{2m}\frac{r^2}{t^2}\right)|2k+1\rangle_1\right)|2m\rangle_2 +$$
$$t \sum_{m=0}^{\infty} \frac{r^{2m}\sqrt{(2m+1)!}}{(2m)!} \left(\sum_{k=0}^{\infty} b_{2(k+m)} t^{2k} \sqrt{\frac{(2(k+m))!}{(2k)!}} \left(1 - \frac{2k}{2m+1}\frac{r^2}{t^2}\right)|2k\rangle_1\right)|2m+1\rangle_2. \quad (10)$$

Since the parity of a single-photon state is transparently odd, registration of an even number of photons $n = 2m$ in mode 2 outputs in mode 1 a state of odd parity of the form

$$|\Psi_{2m}^{(1)}\rangle = K_{2m} \sum_{k=0}^{\infty} b_{2(k+m)} t^{2k} \sqrt{\frac{(2(k+m))!}{(2k+1)!}} \left(1 - \frac{2k+1}{2m}\frac{r^2}{t^2}\right)|2k+1\rangle, \quad (11)$$

where $K_{2m}$ is the normalization factor, which occurs with the probability

$$P_{2m}^{(1)} = \frac{2mt^4(1-t^2)^{2m-1}}{(2m-1)!K_{2m}^2}. \quad (12)$$

with the exception of $n = 0$, while in Eq. (11) the superindex "(1)" indicates that the singe-photon state $|1\rangle$ is inputted into mode 2. Otherwise, if an odd number of photons $n = 2m + 1$ is detected in mode 2, the output state of mode 1 has an even parity of the form

$$|\Psi_{2m+1}^{(1)}\rangle = K_{2m+1} \sum_{k=0}^{\infty} b_{2(k+m)} t^{2k} \sqrt{\frac{(2(k+m))!}{(2k)!}} \left(1 - \frac{2k}{2m+1}\frac{r^2}{t^2}\right)|2k\rangle, \quad (13)$$

with $K_{2m+1}$ the corresponding normalization factor and the probability of this event is

$$P_{2m+1}^{(1)} = \frac{(2m+1)t^2(1-t^2)^{2m}}{(2m)!K_{2m+1}^2}. \quad (14)$$

Of course, $\sum_{m=0}^{\infty}\left(P_{2m}^{(1)} + P_{2m+1}^{(1)}\right) = 1$ as should be. The proximity of the output states (11) and (13) to the odd (4) and even (3) SCSs, respectively, is characterized by the fidelity $F_{2m}^{(1)} = \left|\langle SCS_-|\Psi_{2m}^{(1)}\rangle\right|^2$ and $F_{2m+1}^{(1)} = \left|\langle SCS_+|\Psi_{2m+1}^{(1)}\rangle\right|^2$, respectively, which is completely determined by the set of experimental initial parameters $(s,t)$, the size $\beta$ of the target SCS and the measurement outcome $n \in \{2m, 2m + 1\}$. We numerically plot the dependences of maximum values of the fidelities $F_{2m+1}^{(1)}(\beta)$ and $F_{2m}^{(1)}(\beta)$ in Fig. 4a and Fig. 4b, respectively. Similar to Fig. 2a and Fig. 2b, both $F_{2m+1}^{(1)}(\beta)$ and $F_{2m}^{(1)}(\beta)$ increase with the detected number of photons for a given $\beta$, but decrease with increasing $\beta$ for a given measurement outcome $n \in \{2m, 2m + 1\}$. It is noticed here that the value of amplitude $\beta_{0.99}^{(1)}(n)$ from which the maximized fidelity starts to fall down below 0.99 is larger than the corresponding value $\beta_{0.99}^{(0)}(n)$ in the case when the vacuum state $|0\rangle$ is inputted into mode 2. For example, from Figs. 4a and 4b in comparison with Figs. 2a and 2b, one sees that $\beta_{0.99}^{(1)}(31) \simeq 4.2 > \beta_{0.99}^{(0)}(31) \simeq 3.2$ and $\beta_{0.99}^{(1)}(30) \simeq 4.1 > \beta_{0.99}^{(0)}(30) \simeq 3.1$. This means that the use of a single photon as an input to mode 2 can generate, with high enough fidelity, SCSs of bigger size (i.e., larger value of $\beta$) compared to the case of inputting the vacuum to mode 2. It is also interesting to note that in contrast to the previous consideration with the vacuum state inputted into mode 2, now the probabilities $P_{2m+1}^{(1)}(\beta)$ and $P_{2m}^{(1)}(\beta)$ show up as smooth functions of $\beta$ for each given outcome $n \in \{2m, 2m + 1\}$ without any optimization procedure, as seen in Fig. 4c or Fig. 4d, respectively. In part, this is due to the fact that the fidelities gradually converge to one point of maximal fidelity as shown in Fig. 3b (i.e., there is no long stripe of the fidelities $\geq 0.99$). Yet, the dependence of the probabilities $P_n^{(1)}(\beta)$ on $\beta$ and $n$ is opposite to that of the fidelities $F_n^{(1)}(\beta)$. Namely, as it follows from Figs. 4a, 4b, 4c and 4d, $F_n^{(1)}(\beta)$ increase but $P_n^{(1)}(\beta)$ decrease with increasing $n$ for a given $\beta$, while $F_n^{(1)}(\beta)$ decrease but $P_n^{(1)}(\beta)$ increase with increasing $\beta$ for a given $n$. The plots in Figs. 4e, 4g, 4h and 4i show the dependence of $s, t$ on $\beta$ which provide the fidelity maximum and corresponding success probabilities. So, the plots in Figs. 4e and 4h display values of the parameters $s$ and $t$ under which the plots in Figs. 4a and 4c are obtained. The plots in Figs. 4g and 4i show values of the parameters $s$ and $t$ under which the plots in Figs. 4b and 4d are constructed. There are domains of $\beta$ in which the parameter $s$ ($t$) changes in an abrupt manner as visual from Fig. 4h (Fig. 4i) which is due to the fact that the maximum value of fidelity disappears in one range of values $(s, t)$ and appears in another. Another explanation maybe as follows: it may be due to a large step of varying $\beta$. The fact is that the calculations were carried out not at every point of $\beta$ but only at certain points $\beta_i$ with some step $\delta\beta$ ($\beta_{i+1} = \beta_i + \delta\beta$), which was reflected in a sharp (not smooth) change in some points of $s, t$. Finally, why the values $s, t$ can change so dramatically at the points, maybe it is due to action of both reasons: transition to another region and not so small step $\delta\beta$).

Conditioned on the measurement outcomes, the subscripts of the output states amplitudes are shifted forward by either $m$ ($k \to k + m$) (Eqs. (6,11,13)) or $m + 1$ ($k \to k + m + 1$) (Eq. (8)). This displaces the original SMSV distribution (1) towards Fock states with larger photon numbers, finally becoming uniform distribution, that is $b_{2k} > b_{2(k+1)}$ $\forall k$ but $b_{2k} \sim b_{2(k+1)} \approx 0$ for $k \gg 1$. In addition, each of amplitudes $b_{2(k+m)}$ ($b_{2(k+m+1)}$) receives an extra factor that may amplify them. If one chooses the values of $(s, t)$ in an appropriate way, then the photon number distributions of the conditioned state and the target even/odd SCSs can coincide with fidelity $\geq 0.99$ despite small discrepancy between generated and target probabilities (see Fig. 5). The maximum discrepancy in the probabilities $d_n$, with subscript $n$ indicating on Fock state $|n\rangle$, is $d_{10} = 0.032906$ (top left plot), $d_{11} = 0.031252$ (top right plot), $d_{18} = 0.023296$ (lower left plot) and $d_{17} = 0.05161$ (lower right plot). Discrepancy can affect the fidelity of the transformation of the coherent states on balanced BS: $BS_{12}(|\pm\beta\rangle_1|\pm\beta\rangle_2) = |\pm\sqrt{2}\beta\rangle_1|0\rangle_2$ and $BS_{12}(|\pm\beta\rangle_1|\mp\beta\rangle_2) = |0\rangle_1|\mp\sqrt{2}\beta\rangle_2$, but can be significantly reduced with choice of the experimental parameters providing fidelity very close to 1.

So far we have considered the perfect PNR detection when the quantum efficiency $\eta$ is 1. In reality, PNR detectors are imperfect whose efficiency, though can be very close to 1, remains less than 1, i.e., $\eta < 1$ [28,29]. We model the imperfect detector by placing the fictitious beam splitter of transmissivity $\eta$ before the perfect detector which is responsible for the loss of some of the unregistered photons to derive the positive-operator values measure (POVM) element of the PNR detector with imperfect detection efficiency $\eta < 1$. For example, for $n = 2m$: $\Pi_{2m}(\eta) = \sum_{x=0}^{\infty} C_{2(m+x)}^{2m} \eta^{2m}(1-\eta)^{2x}|2(m+x)\rangle\langle 2(m+x)| + \sum_{x=0}^{\infty} C_{2(m+x)+1}^{2m} \eta^{2m}(1-\eta)^{2x+1}|2(m+x)+1\rangle\langle 2(m+x)+1|$. Finally, we can compute the fidelity between $\rho_{2m}^{(0)} = tr_2\left(\rho^{(0)} \Pi_{2m}(\eta)\right)$, where $tr_2$ stands for trace operation in second mode and $\rho^{(0)}$ is the density matrix, conditioned by measurement of $2m$ photons and target states

$$Fid_{2m}^{(0)} = tr\left(\rho_{2m}^{(0)}(|SCS_+\rangle\langle SCS_+|)\right) = \frac{\sum_{x=0}^{\infty} \frac{(1-\eta)^{2x}(1-t^2)^{2x}}{(2x)! L_{2(m+x)}^2} \left|\langle SCS_+|\Psi_{2(m+x)}^{(0)}\rangle\right|^2}{\sum_{x=0}^{\infty} \frac{(1-\eta)^{2x}(1-t^2)^{2x}}{(2x)! L_{2(m+x)}^2} \left(1 + (1-\eta)\frac{t^2(1-t^2) L_{2(m+x)}^2}{(2x+1) L_{2(m+x)+1}^2}\right)}. \quad (15)$$

where $tr$ for trace operation in first mode. Using the Eq. (15), for $\eta$ such that $1 - \eta \ll 1$, one can decompose the fidelity over PNR detector inefficiency $1 - \eta$: $Fid_{2m}^{(0)}(\eta) = Fid_{2m}^{(0)}(\eta = 1)\left(1 - (1-\eta)g_{2m,1}^{(0)} + (1-\eta)^2 g_{2m,2}^{(0)} + \cdots\right)$, where $g_{2m,1}^{(0)} = t^2(1-t^2) L_{2m}^2 / L_{2m+1}^2$, $g_{2m,2}^{(0)} = (1-t^2)^2 L_{2m}^2 / 2L_{2(m+1)}^2 \left(2t^4 L_{2m}^2/L_{2m+1}^2 + \left|\langle SCS_+|\Psi_{2(m+1)}^{(e)}\rangle\right|^2 / \left|\langle SCS_+|\Psi_{2m}^{(e)}\rangle\right|^2 - 1\right)$. Making use of the relation $b_{2(k+m+1)} = \tanh r \sqrt{(k+m+0.5)/(k+m+1)}\, b_{2(k+m)}$, one can evaluate the ratio $L_{2m}^2/L_{2m+1}^{(e)2} < \tanh r^2(\langle n \rangle + (m+1)^2)$, that enables to construct the lower bound (LB) restricting the fidelity in the case of imperfect photon number detection

$$Fid_{2m}^{(0)}(\eta) > Fid_{2m}^{(0)}(\eta = 1)\left(1 - (1-\eta)t^2(1-t^2)\tanh r^2(\langle n \rangle + (m+1)^2)\right), \quad (16)$$

with $\langle n \rangle$ the average number of photons in the state $|\Psi_{2m}^{(0)}\rangle$. The inequality (16) is valid in the case of at least $t > 0.4$ to provide $g_{2m,1}^{(0)} \gg g_{2m,2}^{(0)}$, otherwise, the contribution of $g_{2m,2}^{(0)}$ will be comparable with one of $g_{2m,1}^{(0)}$. It should be noted that the range of values $t < 0.4$ is incompatible with the generation of high-amplitude ($\beta \geq 2$) even SCS. Similar expression for the even SCS (13) fidelity can be derived

$$Fid_{2m+1}^{(1)}(\eta) > Fid_{2m+1}^{(1)}(\eta = 1)\left(1 - (1-\eta)t^2(1-t^2)\tanh r^2\left(\frac{2(m+1)}{2m+1}\right)^2 (\langle n \rangle + 4m(1+m))\right), \quad (17)$$

where $\langle n \rangle$ is average number of photons in the state $|\Psi_{2m+1}^{(1)}\rangle$ and $g_{2m+1,1}^{(1)} \gg g_{2m+1,2}^{(1)}$ in the case of $t > 0.4$.

The LB of the even SCS (16) is proportional to $m^2$ ($m = n/2$ is half of the detected photon number) and involves average number of the photons $\langle n \rangle$. There are two strategies for realizing even SCSs by imperfect PNR detection. The first strategy is based on the use of high-transmission BS, since the LB tends to disappear in the limit of $t \to 1$ (Fig. 6) when only a small fraction of the input photons can be deflected into the measurement mode (visually, this can be explained using Fig.3a if we trace the curved contour ending at point $t = 1$) and the fidelity of the output state is almost the same as ideal $\left(Fid_{2m}^{(0)}(\eta) \cong Fid_{2m}^{(0)}(\eta = 1), Fid_{2m+1}^{(1)}(\eta) \cong Fid_{2m+1}^{(1)}(\eta = 1)\right)$. But a sufficiently high fidelity in Fig. 6 is accompanied with an extremely low success probability of order of $10^{-18}$, which is far from practical needs. Such generation events become quite rare as the probability to detect $n$ photons has the order $r^n$. Another strategy can be associated with a decrease in the number $m$ and the choice of such values $(s, t)$ that would provide a smaller value of $\langle n \rangle$ estimated as $\langle n \rangle \approx |\beta|^2$ with $\beta$ being an amplitude of the even SCS. This strategy does not use high-transmission BS not to critically reduce the success probability and keep it in an acceptable level, while it allows to compensate contribution of $m$ and $\langle n \rangle$ terms in the LB in Eq. (16). It can be more practical, since it also reduces the requirements to PNR detector to distinguish less numbers of photons, for example, $10$ or $12$ photons instead of $30$. To choose such $(s,t)$ one should descend along the bent line into the region of larger values of $t$ and smaller values of $s$ (see Fig. 3a)). It allows obtaining the fidelity of even SCS of $\beta = 2.5$ at least greater than $0.96$ with acceptable success probability of order $10^{-7}$ when $10$ photons are detected by PNR detector with quantum efficiency $\eta = 0.98$ [29]. The LB of the $Fid^{(1)}_{2m+1}(\eta)$ in Eq. (17) is smaller than the LB of the $Fid^{(0)}_{2m}(\eta)$ in Eq.(16) because $4m > m+1$ for integers $m > 1$. Nevertheless, calculations show that it is possible to realize an even SCS of the amplitude $\beta = 2.8$ with fidelity greater than $0.96$ and with an acceptable success probability by extracting $11$ photons by PNR with $\eta = 0.98$ [29].

Decomposing the odd fidelities over small parameter $1 - \eta$, one gets LB

$$Fid^{(0)}_{2m+1}(\eta) > Fid^{(0)}_{2m+1}(\eta = 1)\left(1 - (1-\eta)\frac{(1-t^2)}{t^2}\langle n \rangle\right), \tag{18}$$

for state (8) and

$$Fid^{(1)}_{2m}(\eta) > Fid^{(1)}_{2m}(\eta = 1)\left(1 - (1-\eta)\frac{(1-t^2)}{t^2}\left(\frac{2m+1}{2m}\right)^2 \langle n \rangle\right), \tag{19}$$

for state (11). The inequalities (18) and (19) are valid for $t > 0.4$ to ensure the predominance of contribution of the first order in $(1-\eta)$ over those of all the higher orders. As in the case of generating even SCS, the range of values $t < 0.4$ does not allow generating odd SCS of larger amplitude ($\beta \geq 2$). The LB can take almost zero value in the case of high transmission BS ($t \to 1$) and $Fid^{(0)}_{2m+1}(\eta) \cong Fid^{(0)}_{2m+1}(\eta = 1)$ but such a choice may not be particularly successful as it may lead to a decrease in the probability as well as a departure from ideal fidelity for odd SCS (see the point of maximum fidelity in Fig. 3b). As can be seen from the expressions (18) and (19), the LBs do not include the terms proportional to either $(m+1)^2$ as in (16) or $4m(1+m)$ as in (17), which increases the chances of generating odd SCS with greater fidelity. BS with transmittance $t > 1/\sqrt{2}$ to compensate multiplier $\langle n \rangle$ in Eqs. (18) and (19) can be in area of interest. Analysis of the data shows that the values of the BS transparency can be chosen in the range $1/\sqrt{2} < t < 0.9$ to provide acceptable generation rate. To reduce $\langle n \rangle$ one can also refuse to extract a large number of photons of either $30$ or $31$ and limit to smaller number of detected photons by inefficient PNR detector. So, in the case of extracting $11$ photons by PNR detector with efficiency $\eta = 0.98$, an odd SCS of amplitude $\beta = 2.5$ can be generated with fidelity of about $0.97$. Analysis of data shows that high-fidelity of about $0.97$ and large-amplitude ($\beta = 3$) odd SCS $|\Psi^{(1)}_{10}\rangle$ in Eq. (11) can be generated with relatively appropriate success probability of order $10^{-7}$ by subtracting $10$ photons by inefficient PNR detector with quantum efficiency $\eta = 0.98$.

Ability to generate large-amplitude and high-fidelity coherent superpositions is important for scalable continuous-variables quantum computation and quantum information processing but the preparation of the qualitative states remains a challenging task. We have proposed and analyzed a simple and efficient way to generate large-amplitude even/odd SCSs with high fidelity and an acceptable for practical use generation rate using irreducible number of the optical elements. In our scheme, even/odd SCSs of amplitude $\beta = 4.2$ with fidelity higher than $0.99$ can be generated with reasonable success probability $\sim 10^{-7} \div 10^{-2}$ in the case of perfect photon number resolving detection. Such an ideal even/odd SCS source would have all three excellent characteristics: size, fidelity and generation rate and be ready for quantum optical computing [7,11].

Original SMSV state as one of certain parity is used for shaping even/odd SCSs. Using SMSV makes sense since extracting any number of photons in an indistinguishable manner preserves the parity of the output state, leaving output either even or odd. The use of undefined parity states can also make a sense, provided that an additional displacement operator will be required. Indeed, even/odd coherent superpositions become uncertain in displaced number state base [26,27]. The generation mechanism is based on the redistribution of the photon states by eliminating the contributions of the vacuum and the lower-photon states, which have a larger weight in the initial SMSV distribution and increasing the contribution of the multiphoton states corresponding to $n \sim |\beta|^2$. It can also be argued that the use of additional non-Gaussian resource (input single photon) makes it possible to improve such a non-classical light source by at least two indicators (size of the generated superposition and its fidelity), while leaving the third indicator (generation rate) within the acceptable range.

In the case of an ideal PNR detector ($\eta = 1$), the relationship between the size of the generated coherent superposition, its fidelity, and the generation rate remains relatively ideal. Such event-ready coherent superposition resource may become ideal for further CV quantum information processing. But using an imperfect detector with $\eta < 1$ spoils this harmonious combination between the three characteristics. In general, two characteristics (size and fidelity) can be kept at the previous level by, for example, the use of an highly transmission beam splitter, but the SCSs generation rate can be significantly reduced in order of magnitude by $10^{-12}$ since only an insignificant part of the photons can be trapped to measurement mode. To keep the third characteristic at a relatively proper level, it is required to use beam splitter with transmittance coefficient in a certain range $1/\sqrt{2} < t < 0.9$ and to reduce the number of extracted photons to either 10 or 11 instead of either 30 or 31. This ratherish reduces the first two characteristics (size and fidelity), but allows us to circumvent a drastic decrease of the generation rate thereby maintaining concord between all three characteristics by measurement by inefficient PNR detector. Interestingly, the effect of the squeezing amplitude on the characteristics of a nonclassical light source is less noticeable as compared with transmittance of the beam splitter. Increasing the squeezing amplitude may not improve the source performance although reduces the weight of vacuum and two-photon states. The beam splitter parameter is more important to maintain the source parameters at an acceptable level. In general, the technology of extracting a large number of photons from the SMSV is promising for SCS generation and relatively easily feasible in the presence of a highly efficient PNR detector. The generation rate would be improved further by the progress of the related technologies, for example, by improving quality of PNR detection or by cascading generation when detecting a large number of photons is divided into detecting a smaller number of photons following each other or use another input CV state of definite parity that requires a separate study.

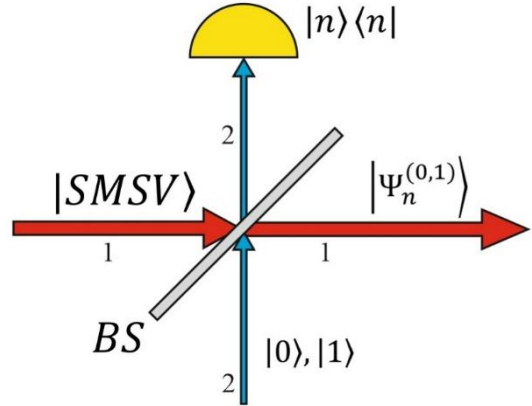

**Fig.1** The optical scheme used to shape either even or odd Schrödinger cat states (SCSs). A single-mode squeezed vacuum state ($|SMSV\rangle$) is inputted into one path (mode 1) of a beam splitter (BS), while either the vacuum state ($|0\rangle$) or a single-photon state ($|1\rangle$) enters the other path (mode 2). Conditioned on the number $n$ of photons detected in the output of mode 2, the output of mode 1 $|\Psi_n^{(0,1)}\rangle$ may be shaped, by choosing proper transmittance coefficient of the BS and the squeezing parameter of the $|SMSV\rangle$, to be a desired bright SCS, either $|SCS_+\rangle$ or $|SCS_-\rangle$, with high fidelity.

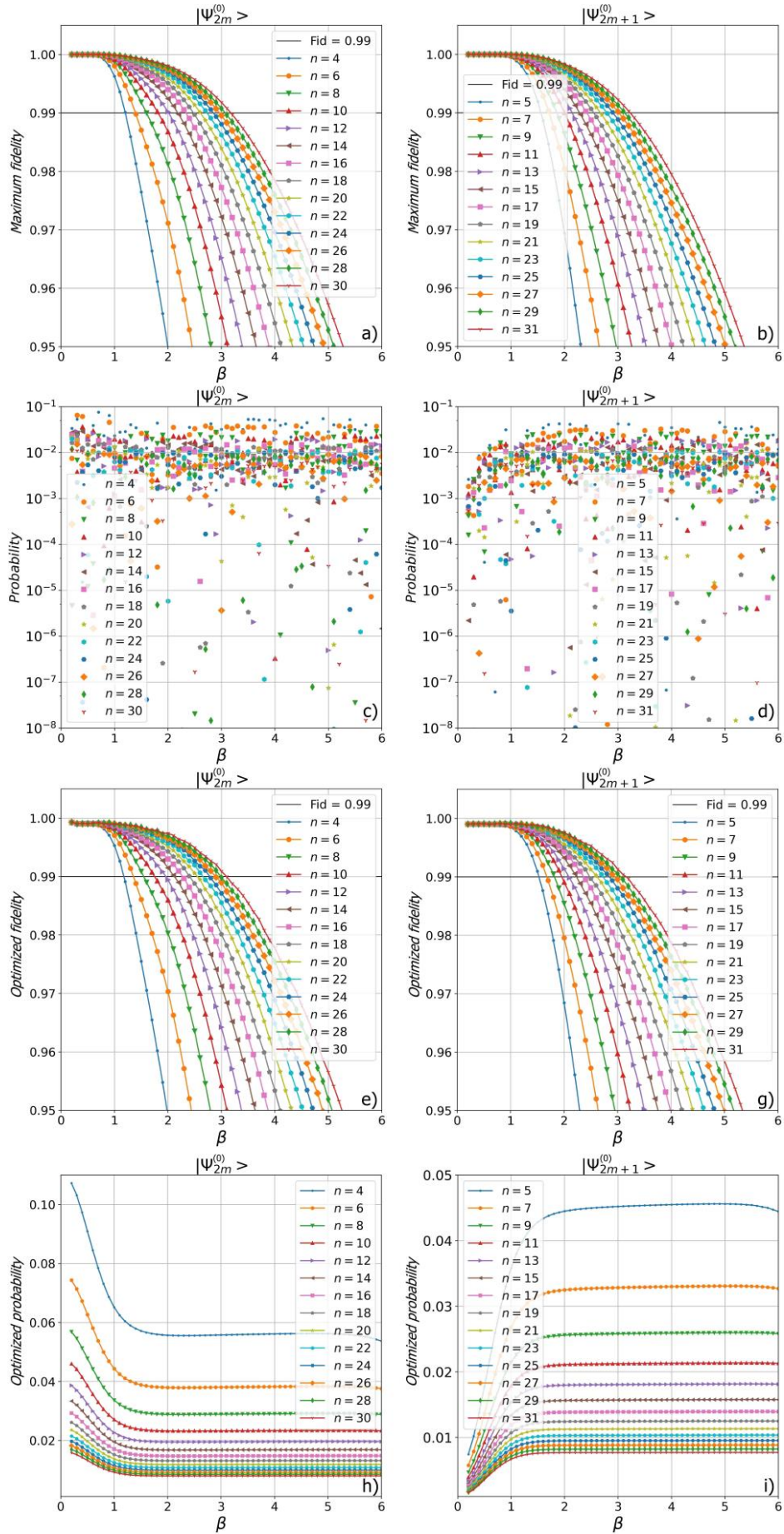

**Fig.2** Dependencies of the maximum values of fidelities $F_n^{(0)}$ and corresponding probabilities $P_n^{(0)}$ on $\beta$ and $n$ when a,c) $n=2m$ and c,d) $n=2m+1$. In a,b) the values of $s,t$ are chosen to maximize the fidelities. With so chosen values of $s,t$ the probabilities are irregularly scattered, as shown in c,d). The quantities in e,g,h,i) are optimized so that the fidelity maintains good enough and the probability becomes highest possible (see text).

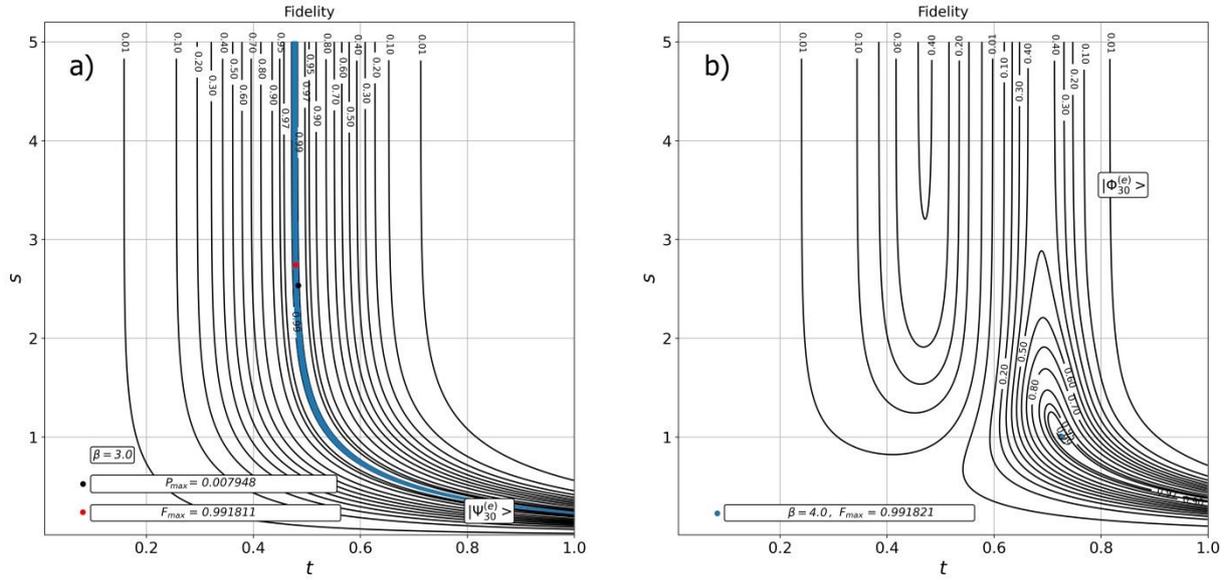

**Fig.3** Isolines versus experimental parameters $(s,t)$: a) for the scheme without an initial single photon and (b) for the scheme with initial single photon. A curved isoline 0.99 (dark blue bent line) going from high values of $s$ to small values of $s$ in a high-transmission beam splitter is observed (a). This isoline explains the swarm distribution in Figs. 2c and 2d. To obtain a regular behavior of the success probability in Figs. 2h and 2i, it is necessary to find $(s,t)_{Prob}$ (black point) on this isoline stripe providing $P_{max}$ unlike $(s,t)_{Fid}$ (red point) providing $P_{max}$ and $(s,t)_{Prob} \neq (s,t)_{Fid}$. The isolines in (b) converge to one point (blue point), which ensures the initially regular behavior of the success probability in Figs. 2c and 2d.

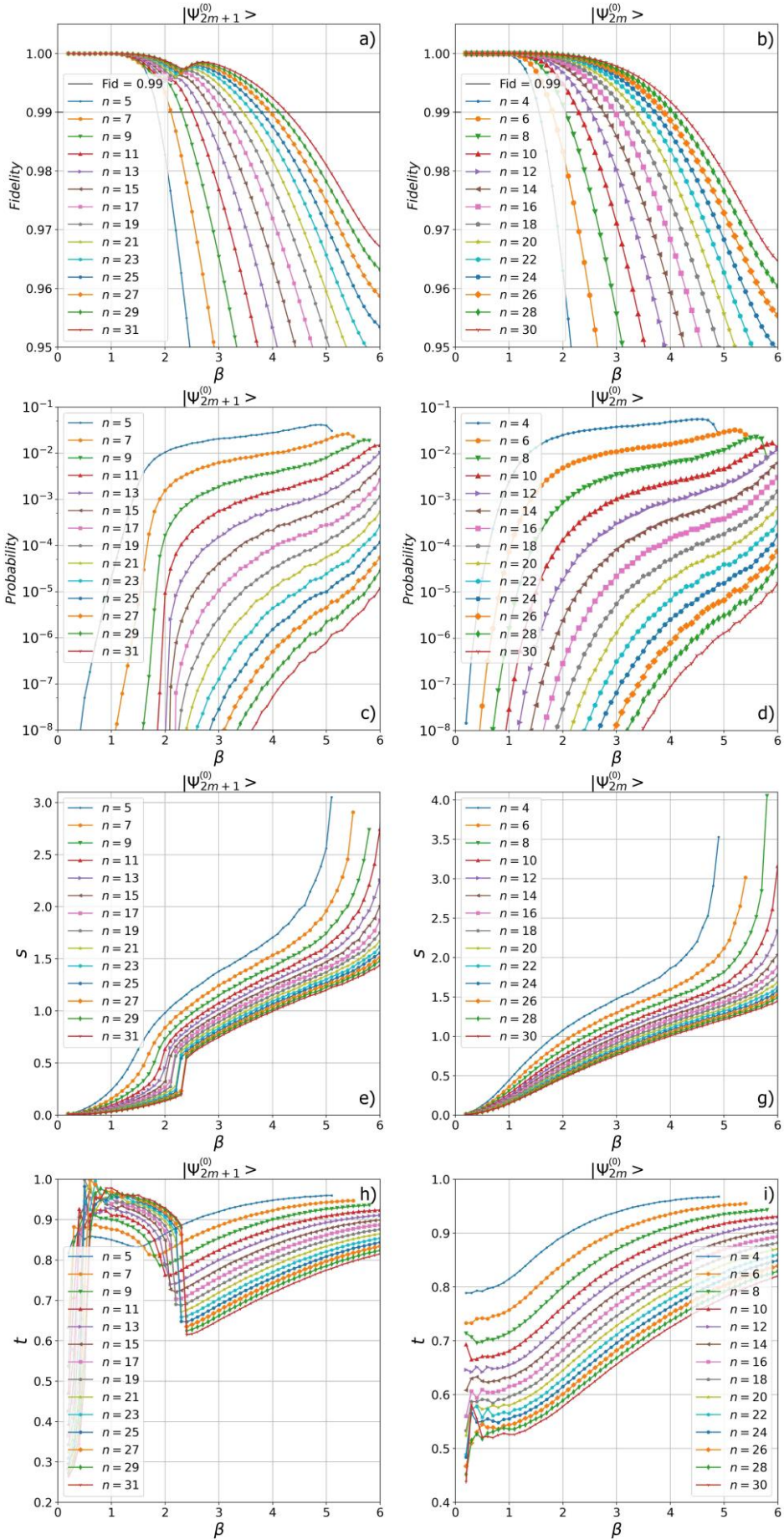

**Fig.4** Dependencies of the maximum values of fidelities $F_n^{(1)}$ and corresponding probabilities $P_n^{(1)}$ on $\beta$ and $n$ when a,c) $n = 2m+1$ and c,d) $n = 2m$. The dependences of $s$ e) and $t$ h) on $\beta$ provide the fidelities and probabilities in a,c), while the dependences of $s$ g) and $t$ i) on $\beta$ provide the fidelities and probabilities in b,d). Sharp changes in $s$ and $t$ (and partly in fidelity) are associated with the transition of the maximum values of the fidelity from one range of parameters $(s, t)$ to another.

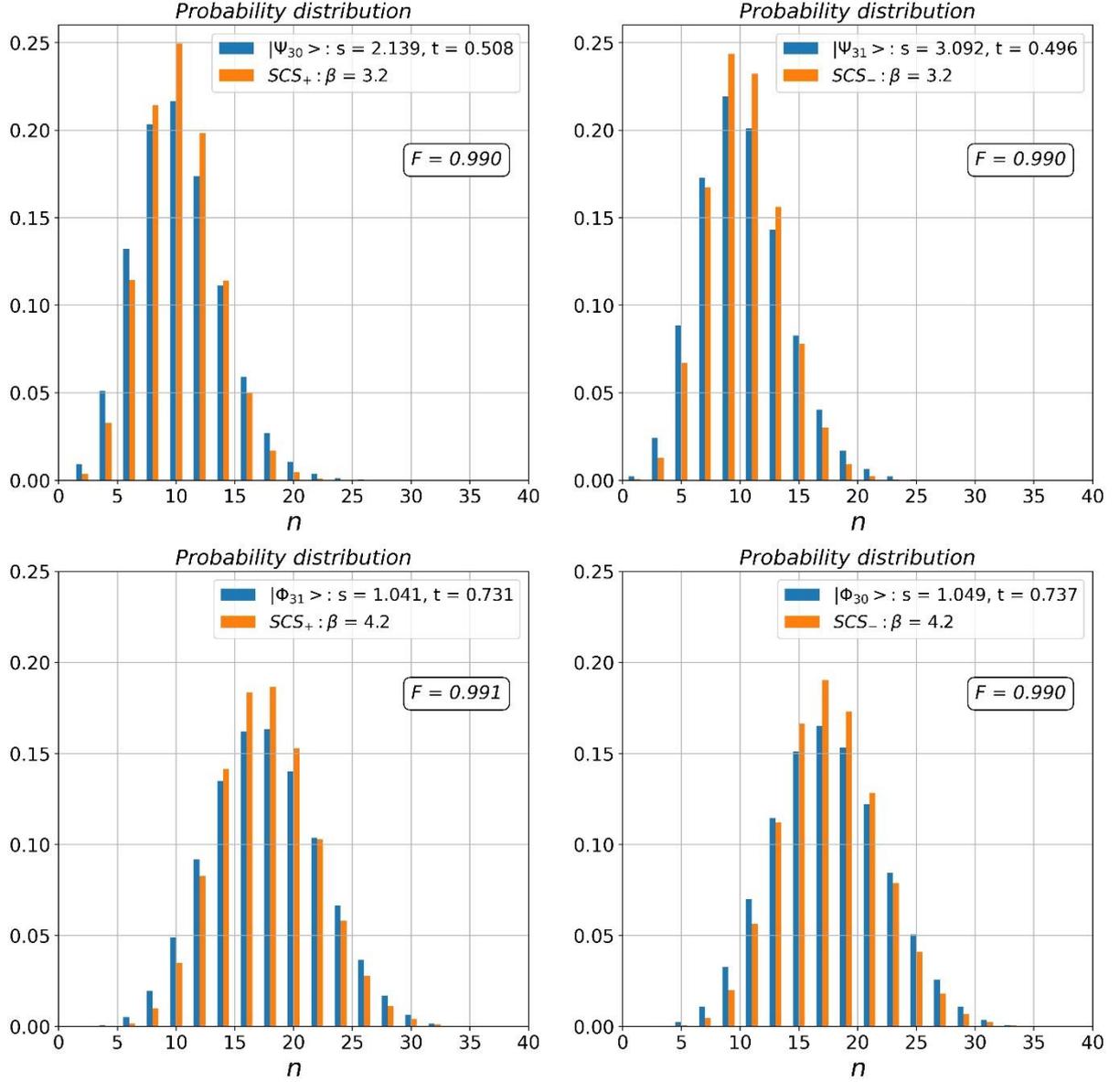

**Fig.5** Comparison of Fock state distributions in conditional and target even/odd SCSs. The maximum calculated discrepancy in the probabilities is $d_{10} = 0.032906$ (difference of probabilities in state $|10\rangle$ for top left distribution), which is sufficient for the states under consideration to coincide with fidelity larger than or equal to 0.99.

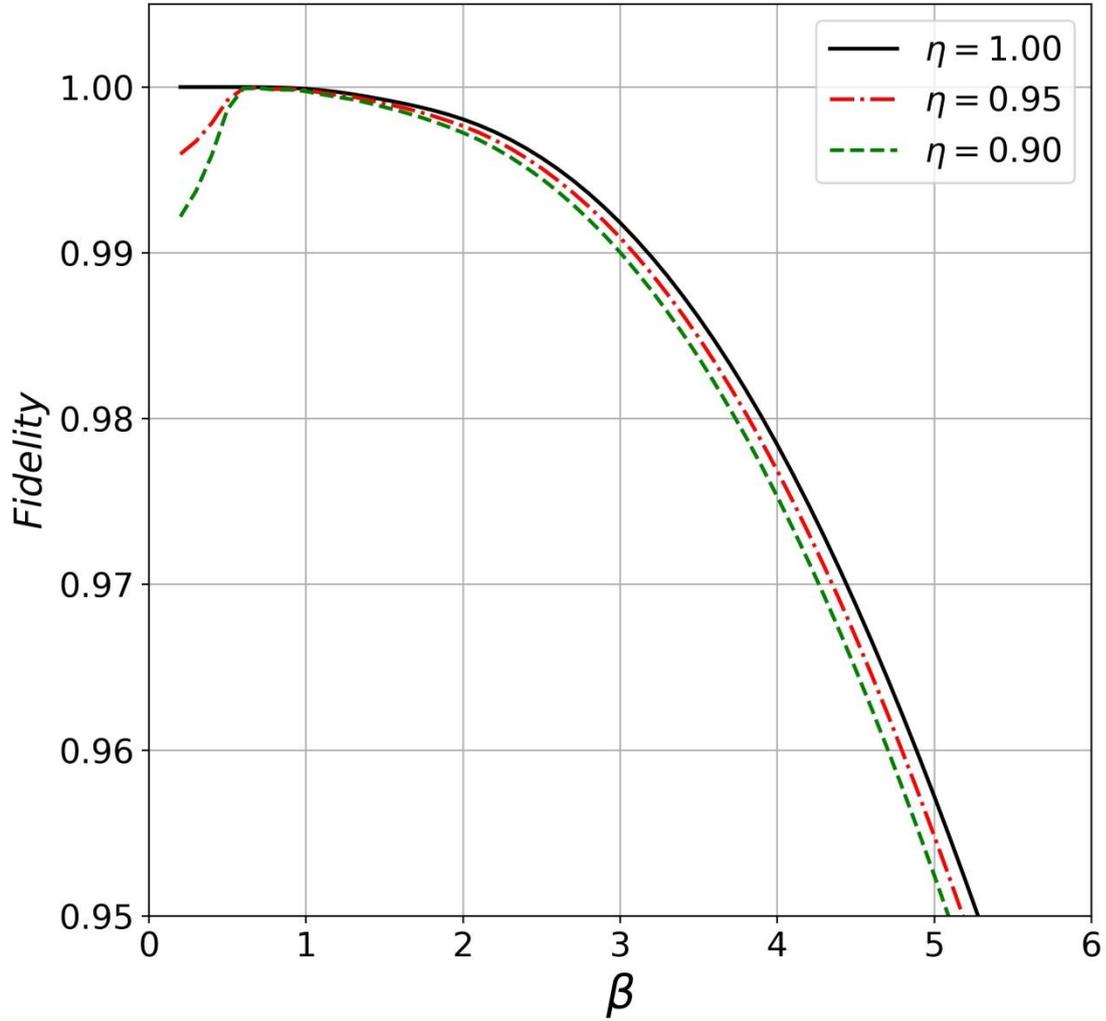

**Fig.6** Dependence of the fidelity on the amplitude of the even SCS generated by the imperfect PNR detector with corresponding quantum efficiency $\eta$ and confirming estimation of the LB in Eq. (16). The dependences are directly calculated from Eq. (15). High transmission beam splitter should be used which drastically reduces the generation rate.